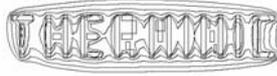

# ADVANCED COMPACT THERMAL MODELING USING VHDL-AMS

*Wasim HABRA[1], Patrick TOUNSI[1,2], Jean-Marie DORKEL[1,2]*

1-LAAS/CNRS - 7, avenue du colonel Roche – 31077 Toulouse cedex 4 – France
2- INSA - 135, avenue de Ranguil – 31077 Toulouse Cedex 4 – France
E-mail: whabra@laas.fr

## ABSTRACT

This paper presents an improved methodology to generate Compact Thermal Models "CTMs" for structures with multi-cooling surfaces and multi-heating sources.

This methodology is based on Star thermal network and makes it possible to have Boundary Conditions Independent "BCI" CTMs.

In order to get BCI models, we use VHDL-AMS modeling language to create variable thermal resistors that change in line with applied boundary conditions.

We use VHDL-AMS, which is IEEE standard modeling language, because it is capable to model any physical phenomena and it is adapted by several manufacturers to simulate devices and systems.

This methodology doesn't need to use any optimizer or additional software and only a few simulations are enough to get BCI model.

The VHDL-AMS blocks that represent these CTMs are sorted by means of the number of heat sources and the number of cooling surfaces. They are saved in a thermal library and can be used for any structure or geometry.

## 1. INTRODUCTION

In the design process of electronic components and systems, it is very important to have an accurate estimate of the thermal behaviour, and because the dissipated power depends on electrical behaviour and the electrical properties depend on the temperature, it is necessary to extend the estimation process to whole electro-thermal problem.

To realize this full electro-thermal simulation, there are several methods which are explained by [1]. As the "relaxation" method has limitations at the compatibility level between simulator and the length of calculation, the direct method is adapted in many research projects which try to extract the compact thermal model of electronic components and try to combine this model with the electrical one.

In this literature, one can find that the existing methodologies have focused on some points of thermal problems and neglected other points because of the complexity of thermal problems and to simplify the generated CTMs.

In this domain, Delphi project has achieved Boundary Conditions Independent (BCI) CTMs valid for different types of electronic packages with a good accuracy [2]. But it becomes very complicated for more than two heat sources. One more, it is cumbersome to generate the model because of the need to create Boundary Conditions set, which can be large, and the use of an optimizer [3].

In [4]&[5] one can find a traditional CTMs and the way to make the thermal coupling between heat sources in static and dynamic modes, but in these methodologies the generated CTMs are BCI only for systems with one cooling surface.

As a result of this discussion we can say that ideal CTMs must have the following points:
- Small number of RC elements.
- Simple model even for multi-sources and multi-cooling structures.
- BCI models.
- 3D behavior of the heat flux as well as the non-linear properties of materials.
- Accuracy in comparison with the 3D thermal analyzing software.

## 2. THE METHODOLOGY FOR MULTICOOLING SURFACES STRUCTURES

As this paper is a continuation of the work which is presented in [6]&[7], we focus here on extracting CTMs for structures with multi cooling surfaces.

### 2.1. Star Model and boundary independence as well

Star thermal network is a very simple representation for thermal problems which allows us to deal easily with dynamic problems and multi-heating sources problems, but the main limitation of this representation is the validity of the thermal model due to changing the boundary conditions (Fig.1).

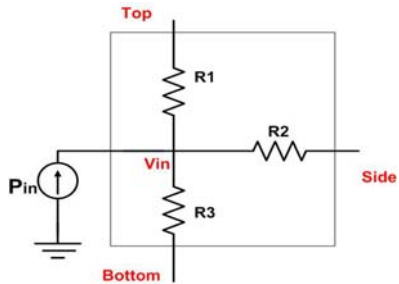

Figure-1: VHDL-AMS block presenting a star CTM.

In our methodology we have adapted the Star network to represent CTMs and we have made some modifications to make it boundary independent.

In [2] we find detailed explanation of the Star model and its limitations compared with Delphi model, but by looking at the value of thermal resistances -which are calculated by (1)- between the junction and each cooling surface, we find that $R_{th-x}$ is minimum when we apply isothermal condition on (x) and adiabatic condition on the other cooling surfaces, on other hand, and we find that $R_{th-x}$ is maximum when we apply isothermal condition on all cooling surfaces (Fig. 2).

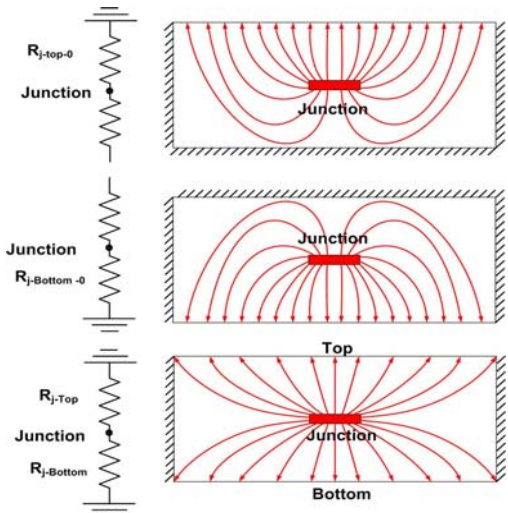

Figure-2: Same network with different thermal resistances according to the applied boundary conditions

$$R_{th-x} = \frac{T_j - T_x}{P_x} \quad (1)$$

Where: -$R_{th}$ is the thermal resistance (k°/W).

-$T_j$, $T_x$ are the temperature of junction and a cooling surface (k°).

-$P_x$ is the power which passes out through the cooling surface (W).

From this observation on the behavior of thermal resistors, we designed variable thermal resistors that vary their values interactively by changing the boundary conditions.

## 2.2. CTM extraction process:

### 2.2.1. Structure description:

In order to explain our methodology, we take the structure presented in fig.3, which is a virtual electronic device with three components. We will assume that only one of them dissipates power, so we will extract the CTM for this source.

The dimension of this device is: (12 x 12 x 3.8)mm where there are several layers (Si, Cu, AL2O3…) and all of them are enclosed in an epoxy package.

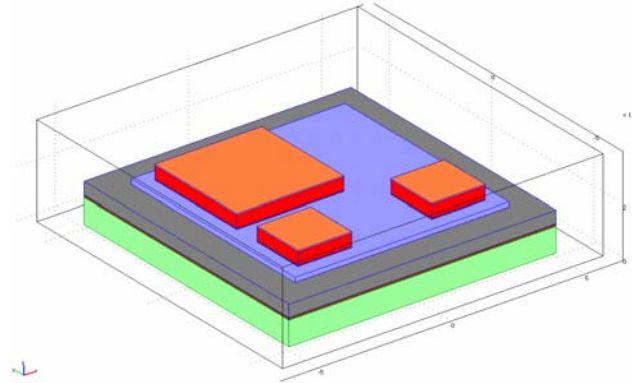

Figure-3: The studied structure in COMSOL®

### 2.2.2. Simulation process:

The first step to generate the CTM is made by defining the cooling surfaces that have the same boundary condition. In our example we assume that all the sides of this device always have the same boundary condition, so we define three parameters (cooling surfaces) which are: Top, Bottom and Side.

The next step is to launch the simulator six times with extreme cooling condition (h=20000 W/m$^2$.k°) applied on the surfaces as follows: Top, Bottom, Side, Top&Bottom, Top&Side and Bottom&Side, and adiabatic condition is applied on the other surfaces for each simulation.

Table-1 presents the results of these six simulations:

|  | $T_j$(C°) | $P_{Top}$(W) | $P_{Bot}$(W) | $P_{Side}$(W) |
|---|---|---|---|---|
| Cooling on Top | 146.4 | 100 | 0 | 0 |
| Cooling on Bot. | 157.35 | 0 | 100 | 0 |
| Cooling on SIDE | 177.32 | 0 | 0 | 100 |
| Cooling on Top & Bot. | 109.33 | 51.75 | 48.16 | 0 |
| Cooling on Top & Side | 122.54 | 57.13 | 0 | 42.53 |
| Cooling on Bot.& Side | 132.87 | 0 | 56.6 | 42.94 |

Table-1: results of 3D detailed model simulation

From this table we calculate the thermal resistances for each case between the junction and each cooling surface by using (1).

Figure-3-a illustrates the changes of the thermal resistance $R_{top}$ versus the changes of thermal flux which pass through cooling surfaces (Bottom and Side), and Figures-3- (b, c) present the same relation for Bottom and Side thermal resistances.

These three plots are the keys to obtain control equations of these three resistors.

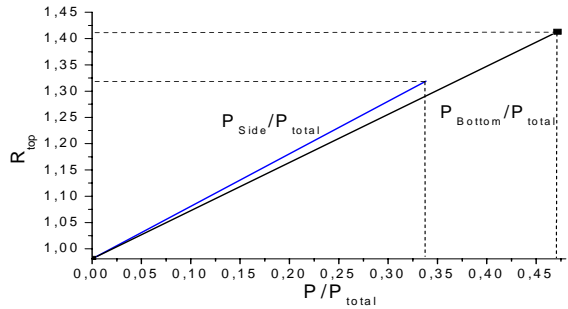

(a)

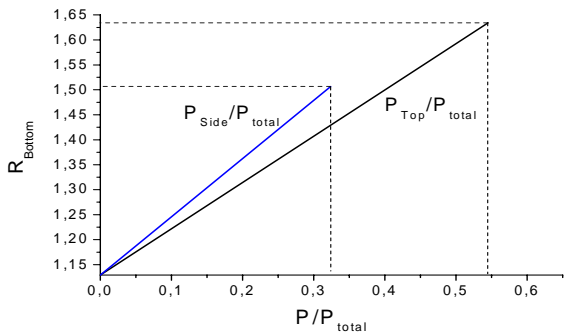

(b)

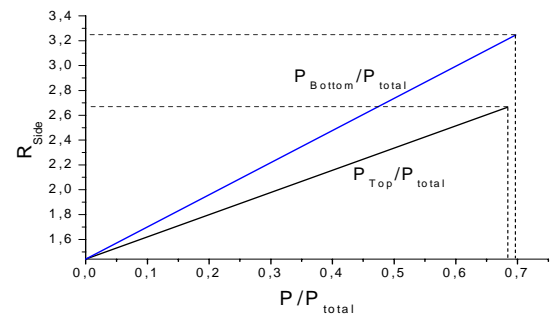

(c)
Figure-4
a: $R_{Top}$ versus $P_{Bottom}/P_{total}$ and $P_{Side}/P_{total}$
b: $R_{Bottom}$ versus $P_{top}/P_{total}$ and $P_{Side}/P_{total}$
c: $R_{Side}$ versus $P_{Bottom}/P_{total}$ and $P_{Top}/P_{total}$

## 2.3. CTM modeling process:

### 2.3.1. Why VHDL-AMS?

To create interactive resistors we have a choice between using a programming language like C++, Basic …etc., or using a modeling language like VHDL-AMS.

We chose VHDL-AMS because it is an IEEE standard and has the ability to model any kind of phenomena (thermal, electrical, mechanical …etc.) and it has been adapted by several manufacturers to simulate systems and devices.

other advantages of using VHDL-AMS include; ease of programming, the fact that it accepts both analog and digital circuit simulation, and some recent VHDL-AMS commercial software even have the ability to simulate projects consisting of VHDL-AMS blocks and PSpice models in the same worksheet (Simplorer®, Systemvision®).

### 2.3.2. The control equation:

In order to simplify the methodology, we have assumed that the changes of thermal resistances in figure-3 are linear.

From the three plots, we can get the following three control equations for the resistors $R_{Top}$, $R_{Bottom}$, $R_{Side}$ consecutively.

$$T_J = \left[ R_{Top\_0} + (\frac{P_{Side}}{P_{Total}}).\alpha_{T-TS} + (\frac{P_{Bottom}}{P_{Total}}).\alpha_{T-TB} \right].P_{Top}$$

$$= \left[ R_{Bottom\_0} + (\frac{P_{Side}}{P_{Total}}).\alpha_{B-BS} + (\frac{P_{Top}}{P_{Total}}).\alpha_{B-TB} \right].P_{Bottom}$$

$$= \left[ R_{Side\_0} + (\frac{P_{Top}}{P_{Total}}).\alpha_{S-TS} + (\frac{P_{Side}}{P_{Total}}).\alpha_{S-SB} \right].P_{Side}$$

where:

- $R_{Top\_min}$, $R_{Bottom\_min}$ and $R_{Side\_min}$ are the thermal resistances that are calculated when we apply an extreme cooling condition on the surface: Top, Bottom and Side consecutively.

- $\alpha_{T-TS}, \alpha_{T-TB}$ : are slopes of curves in Figure-1-a

- $\alpha_{B-BT}, \alpha_{B-BS}$ : are slopes of curves in Figure-1-b

- $\alpha_{S-TS}, \alpha_{S-BS}$ : are slopes of curves in Figure-1-c

- $P_{total} = P_{Top} + P_{Side} + P_{Bottom}$

In VHDL-AMS worksheet, we create a block which inside of we write the equations and the definition.
The compiler of VHDL-AMS solve the four equation where there is only one value for $R_{Top}$, $R_{Bottom}$ and $R_{Side}$ That gives us the same $T_j$.

Each time we change the applied boundary condition on cooling surfaces, the solver search for new values for the three thermal resistances which give the same $T_j$ in the control equations.

## 3. DISCUSSION AND COMPARISION

The presented methodology allows users to obtain simple BCI CTM with a few 3D simulations. Likewise, this methodology is applicable for multi-chips structures and systems and can represent the nonlinear properties of materials.

The required number of 3D simulation to get the CTM is given by:

$$M = n . \sum_{i=1}^{z} i$$

Where:  M is the number of simulation processes.
 $n$ is the number of heat sources
 $z$ is the number of cooling surfaces

The next table presents a comparison between the generated CTM and 3D detailed model (COMSOL®).
We find that the error of our model is generally accepted, but when we apply extreme cooling on the three surfaces at the same time, the error may fall outside of the acceptable range. This error is a result of the assumption that thermal resistances have linear variation when we change the applied boundary conditions.
Reducing this error is possible by introducing a correction factor in the control equations, where this factor depends on the geometry of the studied structure and the value of applied convection.

| Power (W) | Boundary Condition (convection W/m².k°) ||| T (C°) || Error (%) |
|---|---|---|---|---|---|---|
| | Top | Side | Bottom | Simulation 3D | CTM VHDL-AMS | |
| 10 | 10 | 10 | 10 | 2139 | 2133 | 0,22 |
| 10 | 10 | 10 | 100 | 579,2 | 575,43 | 0,42 |
| 10 | 10 | 10 | 1000 | 81,3 | 78,93 | 1,65 |
| 100 | 10 | 10 | 10000 | 195,7 | 181,9 | 2,02 |
| 10 | 100 | 10 | 10 | 578,8 | 574,3 | 0,51 |
| 10 | 1000 | 10 | 10 | 80,8 | 77,49 | 2,31 |
| 100 | 10000 | 10 | 10 | 186,5 | 167,15 | 3 |
| 10 | 10 | 100 | 10 | 489 | 486,66 | 0,39 |
| 10 | 10 | 1000 | 10 | 70,2 | 68,3 | 1,87 |
| 100 | 10 | 10000 | 10 | 207,9 | 199,06 | 1,55 |
| 10 | 100 | 100 | 100 | 225,4 | 220,04 | 1,98 |
| 10 | 100 | 100 | 1000 | 69,5 | 65,9 | 3,21 |
| 100 | 100 | 100 | 10000 | 193,2 | 177,81 | 1,92 |
| 10 | 1000 | 100 | 100 | 69 | 64,76 | 3,85 |
| 100 | 10000 | 100 | 100 | 184,1 | 163,65 | 2,89 |
| 10 | 100 | 1000 | 100 | 62,6 | 60,58 | 2,52 |
| 100 | 100 | 10000 | 100 | 205,5 | 199,38 | 0,47 |
| 100 | 1000 | 1000 | 1000 | 335,3 | 285,53 | 11,85 |
| 100 | 1000 | 1000 | 10000 | 173,7 | 145,39 | 8.1 |
| 100 | 10000 | 1000 | 1000 | 166,4 | 135,9 | 8,34 |
| 100 | 1000 | 10000 | 1000 | 185,6 | 173,05 | 4,29 |
| 100 | 10000 | 10000 | 10000 | 123,1 | 88,97 | 17,97 |

Table-2: comparison between 3D detailed model and our CTM for different boundary conditions.

## 4. PERSPECTIVES

Extending this methodology to include the dynamic problems is very important, and this extension seems possible when we use the Star model network which gives us the simplicity, and when we use VHDL-AMS which gives us the ability to model interactive components.
At the same time it is important to investigate the source of error in the CTM and try to reduce it by extracting a correction factor.

## 6. REFERENCES

[1] H. Gutierrez, C. E. Christoffersen and M. B. Steer, "An integrated environment for the simulation of electrical, thermal and electromagnetic interactions in high-performance integrated circuits," Proc. IEEE 6 th Topical Meeting on Electrical Performance of Electronic Packaging,, Sept. 1999, pp. 217–220.

[2] C.J.M. Lasance, H. Vinke, H. Rosten "Thermal characterization of electronic devices with BCI compact Models" IEEE Trans. On Components, Packaging and Manufacturing Tech. Part A Vol. 18 , Dec. 1995, pp. 723 –729.

[3] H. Rosten C.J.M. Lasance, J.D. Parry, "The world of thermal characterization according to DELPHI- PartI , Part II" IEEE Trans. On Components, Packaging and Manufacturing Tech. Part A Vol. 20 , Dec. 1997, pp. 384 –398.

[4] M. Rencz, G. Farkas, A. Poppe, V. Szekely, B. Coutois. "A Methodology for the Generation of dynamic Compact Models of packages and Heat Sinks from Thermal Transient measurements". IEEE. CPMT/Semi int. Electronics Manufacturing technology symposium. 2003, pp. 117 –123.

[5] M. Rencz, G. Farkas, A. Poppe, V. Szekely, B. Coutois. "An Alogarithm for the direct co-simulation of dynamic compact models of packages with detailed thermal models of boards". IEEE. Electronics packaging technology conference. 2002, pp. 293 –298.

[6] W. Habra, P. Tounsi, J.M. Dorkel. "Improved 3D-nonlinear compact modeling for power components". EuroSime'2005. p.390 –393.

[7] W. Habra, P. Tounsi, J.M. Dorkel. "Transient Compact modeling for multi chips components". THERMINIC'2005.